\documentclass[iop]{emulateapj}

\shorttitle{Dwarf-Dwarf Mergers}
\shortauthors{Deason, Wetzel, Garrison-Kimmel}
\slugcomment{ApJ accepted}

\begin{document}

\title{Satellite Dwarf Galaxies in a Hierarchical Universe: The Prevalence of Dwarf-Dwarf Major Mergers}

\author{Alis Deason\altaffilmark{1}}
\affil{Department of Astronomy and Astrophysics, University of California Santa Cruz, Santa Cruz, CA, USA; alis@ucolick.org}
 
\author{Andrew Wetzel}
\affil{TAPIR, California Institute of Technology, Pasadena, CA, USA}
\affil{Observatories of the Carnegie Institution for Science, Pasadena, CA, USA}

\author{Shea Garrison-Kimmel}
\affil{Center for Cosmology, Department of Physics and Astronomy, University of California, Irvine, CA, USA}

\altaffiltext{1}{Hubble Fellow}

\date{\today}

\begin{abstract}
Mergers are a common phenomenon in hierarchical structure formation,
especially for massive galaxies and clusters, but their importance for
dwarf galaxies in the Local Group remains poorly understood. We
investigate the frequency of major mergers between dwarf galaxies in the
Local Group using the ELVIS suite of cosmological zoom-in
dissipationless simulations of Milky Way- and M31-like host halos. We
find that $\sim 10\%$ of satellite dwarf galaxies with
$M_{\rm star} > 10 ^ {6} M_\odot$ that are within the host virial radius
experienced a major merger of stellar mass ratio closer than 0.1 since
$z = 1$, with a lower fraction for lower mass dwarf galaxies. Recent
merger remnants are biased towards larger radial distance and more
recent virial infall times, because most recent mergers occurred shortly
before crossing within the virial radius of the host halo.
Satellite-satellite mergers also occur within the host halo after virial
infall, catalyzed by the large fraction of dwarf galaxies that fell in
as part of a group. The merger fraction doubles for dwarf galaxies
outside of the host virial radius, so the most distant dwarf galaxies in
the Local Group are the most likely to have experienced a recent major
merger. We discuss the implications of these results on observable dwarf
merger remnants, their star formation histories, the gas content of
mergers, and massive black holes in dwarf galaxies.

\end{abstract}


\bibliographystyle{apj}

\section{Introduction}

Mergers are instrumental in the hierarchical formation of cosmological structure. Galaxy mergers play a crucial role in galaxy evolution, shaping many important properties of galaxies, such as morphological transformation, star formation rates, and quasar activity \citep[for example,][]{toomre72,barnes91,carlberg90}. Moreover, the majority of dark matter accretion into massive halos is in the form of major and minor mergers \citep[e.g.,][]{fakhourima10}.

Less massive halos tend to have accumulated more of their mass at early times, while more massive halos typically had more recent mass growth \citep[e.g.,][]{navarro97,bullock01,wechsler02,Cohn_2005}. Thus, while merger activity continues down to the lower mass scales of dwarf galaxies ($M_{\rm star} \lesssim 10 ^ 9 M_\odot$), the incidence of recent mergers is lower than for higher mass systems \citep[e.g.,][]{wetzel09b,fakhourima10}. However, there are few theoretical investigations into \textit{dwarf-dwarf} galaxy mergers, and observational evidence, in either direction, is scarce.
In particular, the expected merger rates for dwarf galaxies in Local Group-like environments remains poorly known.
Furthermore, while merger rates of halos and subhalos have been studied extensively \citep[e.g,][]{gottlober01,guo08,Wetzel_2008,Angulo_2009,genel09,Stewart_2009,wetzel09a,wetzel09b,fakhouri10}, it is not trivial to relate these to \textit{galaxy} major merger rates because of the non-linear relation between galaxy stellar mass and halo mass \citep[e.g.,][]{behroozi10,moster10,Hopkins_2010}.

Our current view of galaxy formation at the smallest scales is limited to observations of satellites orbiting the host halo of the Milky Way (MW) or M31. Despite the many successes of the Cold Dark Matter (CDM) paradigm on large scales, several discrepancies are uncomfortably apparent on the smallest scales. Perhaps most striking is the dearth of low-mass dwarf galaxies in the Local Group, often termed the ``missing satellites problem''. Simulations predict a wealth of substructure around MW/M31-mass galaxies, although only $\sim 10$ bright ($L > 10^5 L_\odot$) satellites orbit the MW. Recent discoveries have boosted the numbers at the faint end \citep[e.g.,][]{willman05,belokurov06,belokurov07}, but even now the discrepancy can only be alleviated by assuming that the majority of satellites are fainter than the detection limits of current surveys \citep{tollerud08}. Further tension with the CDM model becomes apparent in the internal structure of dwarf galaxies. Observed dwarfs have shallower inner dark matter density profiles than predicted, called the ``core-cusp problem'' \citep{flores94,moore94, gentile04}, and $N$-body simulations predict more dense dark matter subhalos than observed, called the ``too big to fail problem'' \citep{boylan11,boylan12}.

The proximity of the dwarf galaxies in the Local Group allows for detailed scrutiny, but it also means that they can be influenced by the environment of the host galaxies/halos. Various properties of dwarf galaxies, such as gas content \citep[e.g.,][]{grcevich09}, star formation history \citep[e.g.,][]{grebel03,weisz11}, and morphology \citep[e.g.,][]{lisker07} are strongly correlated with the proximity to a large galaxy. Furthermore, satellites can be deformed, disturbed, or tidally stripped via interactions with a large host galaxy \citep[e.g.,][]{mayer01, penarrubia08b}. Indeed, environmental effects such as tidal and ram-pressure stripping are important for baryonic solutions to the ``too big to fail problem'' \citep[e.g.,][]{arraki14, brooks14}. One possible way to disentangle these environmental effects is to study ``isolated'' (beyond the virial radius of the MW or M31) dwarf galaxies in the Local Group \citep[e.g.,][]{kirby14,GarrisonKimmel14b}. However, a large host halo still can have influenced dwarf galaxies at several virial radii, especially for satellites on high-energy orbits that extend beyond the host virial radius \citep[e.g.,][]{sales07,teyssier12,wetzel14}. Seemingly isolated dwarf galaxies could be affected by an encounter with another (dwarf) galaxy, but the importance of this is poorly known at these low masses.

Dwarf galaxies in the Local Group also are excellent laboratories to study star formation history and chemical evolution on the smallest scales. The chemical evolution models used to analyze both spectroscopic and photometric data span a wide range in sophistication and detail \citep[e.g.,][]{marcolini08,revaz09,sawala10,kirby11a,kirby11b}. However, it is generically assumed in these models that the dwarf galaxies are isolated. This may be a valid assumption for many of the dwarf galaxies of the Local Group, though observations of kinematic and chemical peculiarities in, for example, Fornax \citep{battaglia06,deboer12,hendricks14}, Sculptor \citep[e.g.,][]{tolstoy04}, Carina \citep[e.g.,][]{venn12} and And II \citep{ho12} have led to the suggestion that external factors, such as dwarf-dwarf encounters, should be considered more seriously.

At present, it remains unclear both how frequent such dwarf-dwarf interactions are in the Local Group, and how important these events are in the evolutionary history of low-mass galaxies.
The possibility of dwarf-dwarf major mergers generally is ignored, but in this work we aim to address the validity of this assumption, using a suite of cosmological zoom-in dissipationless simulations to characterize the incidence of dwarf-dwarf galaxy mergers in the Local Group. 

The paper is arranged as follows. Section 2 briefly describes the ELVIS suite of simulations and defines major mergers in the context of this work. Section 3 outlines our results, and in Section 4 we discuss the implications for dwarf galaxies in the Local Group. Finally, Section 5 summarizes our main results.

\section{Numerical Methods}

\subsection{ELVIS Simulations}

To study galaxy-galaxy mergers, we use ELVIS (Exploring the Local Volume in Simulations), a suite of zoom-in $N$-body simulations that model the Local Group in a fully cosmological context \citep{GK14}. ELVIS simulates 48 dark matter halos of mass similar to the MW or M31 ($M_{\rm vir} = 1 - 3 \times 10 ^ {12} M_\odot$) within a zoom-in volume of radius $> 4\,R_{\rm vir}$ of each halo (corresponding to $r > 1.4$ Mpc) at $z = 0$. Half of these zoom-in regions contain a pair of halos that resemble the masses, distance, and relative velocity of the MW-M31 pair, while the other half are single isolated mass-matched halos.
In this work, we use both catalogs together and do not distinguish between isolated and paired halos (see \$\ref{sec:merge_fracs}).

ELVIS was run using GADGET-3 and GADGET-2 \cite{Springel_2005} with initial conditions generated using MUSIC \citep{Hahn_2011}, with $\Lambda$CDM cosmology with parameters based on \textit{Wilkinson Microwave Anisotropy Probe} WMAP7 \citep{larson11}: $\sigma_8=0.801$, $\Omega_M = 0.266$, $\Omega_\Lambda = 0.734$, $n_s = 0.963$ and $h = 0.71$.
The zoom-in regions were chosen from a suite of larger simulations each with cubic volume of side length 70.4 Mpc.
Within the zoom-in regions, the particle mass is $1.9 \times 10 ^ 5 M_\odot$ and the Plummer-equivalent force softening is 140 pc (comoving at $z > 9$, physical at $z < 9$).
Three of the isolated halos were run at higher resolution with particle mass $2.4 \times 10 ^ 4 M_\odot$ and force softening of 70 pc, and we use these simulations to check that resolution effects do not significantly affect our results. 
See \citet{GK14} for more details on ELVIS.

\subsection{Finding and tracking (sub)halos}

ELVIS identifies dark matter (sub)halos with the six-dimensional halo finder \textsc{rockstar} \citep{behroozi13a} and constructs merger trees using the \textsc{consistent-trees} algorithm \citep{behroozi13b}.
For each isolated (host) halo that is not a subhalo (within the virial radius of a more massive host halo), we assign a virial mass, $M_{\rm vir}$, and radius, $R_{\rm vir}$, using the evolution of the virial relation from \citet{bryan98} for our $\Lambda$CDM cosmology.
At $z = 0$, this corresponds to an overdensity of $97 (363) \times$ the critical (matter) density of the Universe.

We assign the primary progenitor (main branch) at each snapshot based on the total mass up to and including that snapshot, that is, the main branch contains the most total mass summed from the (sub)halo masses over all preceding snapshots in that branch.
For each (sub)halo, we compute the maximum (peak) mass ever reached by the main branch of a progenitor, $M_{\rm peak}$.
As explored in \citet{GK14}, the (sub)halo catalogs are complete even below the mass limits that we explore in this work ($M_{\rm star} > 10 ^ 3 M_\odot$, corresponding to $M_{\rm peak} \gtrsim 10 ^ 8 M_\odot$, see below).

\subsection{Defining Major Mergers}

To define mergers, at each snapshot of the simulation we follow back all progenitors of each (sub)halo.
For each progenitor that is not the primary progenitor (main branch), we calculate its $M_{\rm peak}$. We then identify the non-primary progenitor that has the highest peak mass, $M_{\rm peak, 2}$, and we define the merger mass ratio as $M_{\rm peak, 2} / M_{\rm peak, 1}$, where $M_{\rm peak, 1}$ is the peak mass of the primary progenitor \textit{before} the snapshot under consideration. Thus, the merger time is when the (sub)halos coalesce fully in phase space such that the finder no longer can distinguish between them. We assume that this timescale is a good proxy for the onset of the \textit{galaxy} merger, as any interaction between the baryonic components of the dwarfs occurs after the majority of the dark matter has been stripped (\citealt{penarrubia08a}). We define mergers with $M_{\rm peak, 2} / M_{\rm peak, 1} > 0.3$ as major mergers, as we justify below.

\subsection{Assigning stellar mass to (sub)halos}

While ELVIS includes only dark matter, our goal is to match dark matter (sub)halos to luminous galaxies.
The relation between stellar mass and (sub)halo dark mass (or circular velocity) for dwarf galaxies is highly uncertain, likely with significant scatter. Nonetheless, we use the relation derived by \citet{GK14} based on ELVIS and observations of dwarf galaxies in the Local Group. They used a modified version of the relation from \citet{behroozi13c}, based on the observed stellar mass function of \citet{baldry12}, which better reproduces dwarf galaxies ($M_{\rm star} < 10 ^ 9 M_\odot$) in the Local Group \citep[Figure 10 in][]{GK14}.
At these mass scales: $M_{\rm star} \propto M ^ {1.92}_{\rm peak}$.

In defining the stellar mass ratios for major mergers in the histories of dwarf galaxies, we assume that the \textit{slope} (but not necessarily normalization) of this relation does not evolve, motivated by the lack of strong evolution measured for slightly more massive galaxies \citep{Wake_2011, Leauthaud_2012, hudson13}, in addition to the lack of observational evidence to suggest otherwise.

We define major mergers as those for which the stellar mass ratios are closer than 0.1.
This broadly corresponds to mass ratios at which the lower mass companion is likely to have significant dynamical effect on the more massive galaxy \citep[e.g.,][]{Hopkins_2010, helmi12, yozin12} and where recent mergers are likely to be observable.
Given the relation between stellar and halo mass, this corresponds to $M_{\rm peak}$ ratios $> 0.3$, as given above.

We note that our adopted threshold for major mergers is dependent on the $M_{\rm star}-M_{\rm peak}$ relation, and, as stated above, this is highly uncertain in this low-mass regime. For example, if we instead adopt a power-law relation with a slope steeper/shallower by $\pm 0.5$dex, then our fiducial $M_{\rm peak}$ ratio criteria for major mergers ($M_{\rm peak, 2} / M_{\rm peak, 1} > 0.3)$ corresponds to stellar mass ratios of $0.05$ and $0.18$, respectively. Thus, steeper $M_{\rm star}-M_{\rm peak}$ relations lead to the inclusion of less significant stellar mass mergers in our analysis, and shallower slopes restrict our criteria to more significant stellar mass mergers. However, it is worth noting that the definition of a ``significant'' merger event itself is not well defined. In the present analysis, we argue that we are using the $M_{\rm star}-M_{\rm peak}$ relation that most closely matches the observations in this low-mass regime, so a stellar mass ratio of $0.1$ is a good estimate for the major merger criteria that we have adopted.

Low-mass galaxies likely have significant stochasticity in their stellar-halo mass relation, though the level of scatter is poorly constrained.
Thus, we also explore the effects of including 0.3 dex log-normal scatter in $M_{\rm star}$ at fixed $M_{\rm peak}$, assuming that this scatter does not correlate between merging galaxies, by applying a 0.42 dex ($\sqrt{2} \times 0.3$) log-normal scatter in the mass \textit{ratio} for mergers, and estimating the median major merger fractions from 100 trials (see Section 3). Note that during this exercise we include events with small $M_{\rm peak}$ ratio, $M_{\rm peak, 2} / M_{\rm peak, 1} > 0.3$, but stellar mass ratio $<0.1$. This takes into account the possibility of ``dark'' mergers (see below), but we note that the inclusion/exclusion of these events only makes a small difference to the quoted fractions.
Including such random scatter boosts the major merger fractions, because it includes mergers between (sub)halos with even more discrepant $M_{\rm peak}$ ratios, which are much more common \citep[e.g.,][]{wetzel09b, fakhouri10}.
Given the power-law relation between stellar-halo mass that we use, including scatter boosts the normalization of merger fractions essentially independent of mass (see Table~\ref{tab:fractions}), but beyond that, the general trends in this paper are unaffected.

We also note the possibility that some of the lowest-mass galaxies may be ``dark'', such that they currently do not contain any (observable) baryons. While these cases would not represent galaxy mergers in a normal sense, a ``dark'' merger still can have a significant impact on a luminous galaxy \citep[e.g.,][]{helmi12}, so these cases remain relevant as merger events.

Finally, we caution that our analysis of dwarf-dwarf mergers is based on dissipationless simulations. The lifetime of subhalos can differ in hydrodynamic simulations, although it is not entirely clear how this will affect subhalo survivability (i.e. more or less survivors). For example,  \cite{guo11} argue for an increased survivability of ``orphan'' galaxies (galaxies without dark matter halos) based on semi-analytic models, whereas \cite{zolotov12} find that subhalos are more efficiently destroyed in hydrodynamic simulations (that include a baryonic disk component) relative to dark matter only simulations.  The required resolution to investigate major mergers at these low mass scales with sufficient statistics is challenging for current state-of-the-art hydrodynamic simulations. Thus, while an investigation with hydrodynamic simulations (at similar resolution) would be favorable, the current analysis offers a useful, and much needed, first step.

\section{Results}

\begin{figure}
\begin{center}
\includegraphics[width=8.5cm, height=6.8cm]{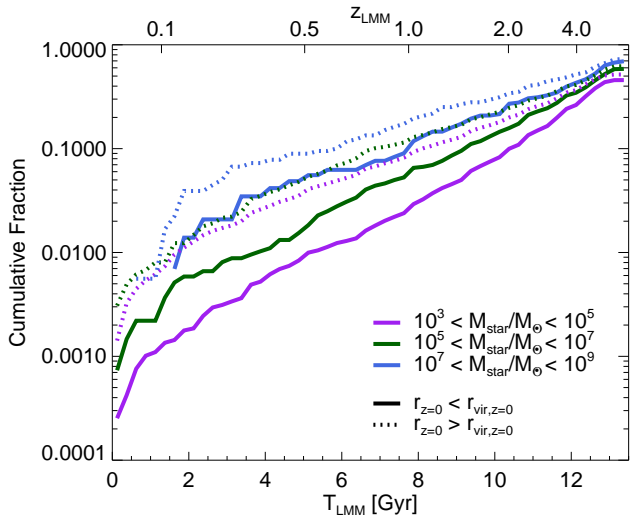}
\caption{\label{fig:tmm} Cumulative fraction of dwarf galaxies that underwent a major merger as a function of time since last major merger, $T_{\rm LMM}$ (or $z_{\rm LMM}$, top axis). Major mergers are those with $M_{\rm star, 2} / M_{\rm star, 1} \gtrsim 0.1$, corresponding to $M_{\rm peak, 2} / M_{\rm peak, 1} > 0.3$. Different colors show different stellar mass bins, while solid/dotted curves show dwarf galaxies that are inside/outside of the host virial radius, $r_{\rm vir}$, at $z = 0$. Major mergers are more common amongst more massive dwarf galaxies and those beyond $r_{\rm vir}$. $45 - 70\%$ of all surviving dwarf galaxies ever experienced a major merger, depending on mass, and $10\%$ of all satellites with $M_{\rm star} >  10 ^ {6} M_\odot$ experienced a major merger since $z = 1$.}
\end{center}
\end{figure}

\subsection{Frequency of Dwarf-Dwarf Mergers}
\label{sec:merge_fracs}
Fig.~\ref{fig:tmm} shows the cumulative fraction of dwarf galaxies that experienced a dwarf-dwarf major merger as a function of time since the last major merger, $T_{\rm LMM}$, or similarly the redshift, $z_{\rm LMM}$. Different colored curves show different mass bins, and solid/dotted curves show dwarf galaxies inside/outside of the host virial radius, $r_{\rm vir}$, at $z = 0$. Table~\ref{tab:fractions} summarizes the relevant merger fractions for different bins of mass and redshift. Dwarf-dwarf mergers were more common at earlier times, and the majority of major mergers occurred prior to $z = 1$, in agreement with previous work \citep[e.g.,][]{klimentowski10}.

The frequency of major mergers depends on galaxy mass. For satellites of higher mass ($M_{\rm star} > 10 ^ 6 M_{\odot}$, corresponding to Fornax, Leo I, or Sculptor in the MW; And I, And II, or And VII in M31), 10\% experienced a major merger since $z = 1$. At lower masses ($M_{\rm star} = 10 ^ {5 - 6} M_\odot$, corresponding to Sextans or Draco) these fractions drop to $\sim 5\%$ , down to $\sim 3\%$ at $M_{\rm star} = 10 ^ {3 - 5} M_\odot$, the mass regime of the ultra-faints. Thus, particularly for the most massive satellites of MW and M31, recent major mergers ($z \lesssim 1$) are not particularly rare. While the redshift baseline of $z < 1$ is somewhat arbitrary, it is motivated by most dwarf galaxies having built the majority of their stellar mass by $z \sim 1$ \citep[e.g.,][]{weisz11}, implying that mergers since $z = 1$ are likely to have observable consequences on a galaxy's evolution.

\begin{deluxetable}{l| c c}
\tablewidth{2.8in} 
\tablecaption{Percentage of dwarf galaxies that underwent a major merger since $z = z_{\rm LMM}$}
\startdata
\multicolumn{3}{c}{\boldmath{$10^7 < M_{\rm star}/M_\odot < 10^9$}}\\
\hline
\\[-0.75em]
 & $r > r_{\rm vir}$ & $r < r_{\rm vir}$ \\
\\[-0.75em]
$z_{\rm LMM} < 0.5$ & 8.9(11.5) & 5.3(6.7) \\
$z_{\rm LMM} < 1.0$ & 18.4(23.7) & 10.4(14.5)\\
$z_{\rm LMM} < 2.0$ & 33.9(43.8)& 26.7(34.2)\\
\multicolumn{3}{c}{\boldmath{$10^5 < M_{\rm star}/M_\odot < 10^7$}}\\
\hline
\\[-0.75em]
 & $r > r_{\rm vir}$ & $r < r_{\rm vir}$ \\
\\[-0.75em]
$z_{\rm LMM} < 0.5$ & 5.1(6.9) & 1.7(2.3)\\
$z_{\rm LMM} < 1.0$ & 12.2(15.9) & 5.9(7.7)\\
$z_{\rm LMM} < 2.0$ & 24.5(32.0)& 15.7(20.8)\\
\multicolumn{3}{c}{\boldmath{$10^3 < M_{\rm star}/M_\odot < 10^5$}}\\
\hline
\\[-0.75em]
 & $r > r_{\rm vir}$ & $r < r_{\rm vir}$ \\
\\[-0.75em]
$z_{\rm LMM} < 0.5$ & 3.9(5.0) & 0.9(1.3)\\
$z_{\rm LMM} < 1.0$ & 8.7(11.3) & 2.7(3.7)\\
$z_{\rm LMM} < 2.0$ & 19.8(25.4)& 9.9(13.2)
\enddata
\tablecomments{\small Values in parentheses are percentages if we assume 0.3 dex scatter in the $M_{\rm star} - M_{\rm peak}$ relation.}
\label{tab:fractions}
\end{deluxetable}
Dwarf galaxies that are beyond the host $r_{\rm vir}$ but still within the volume of the Local Group (distance to nearest host, $r < 1.4$ Mpc) are twice as likely to have experienced a recent major merger (see Table~\ref{tab:fractions}). This increase in merger fractions outside of $r_{\rm vir}$ likely results from the lower relative velocities of galaxies there \citep[e.g.,][]{derijcke04}.

In Table~\ref{tab:fractions}, the values in brackets show merger fractions if we assume a 0.3 dex scatter in our fiducial $M_{\rm star} - M_{\rm peak}$ relation. Using 100 trials, we estimate the major merger fractions and quote the median values. Including scatter increases merger fractions by a factor of $\approx 1.3$, because minor $M_{\rm peak}$-ratio mergers are much more common than major $M_{\rm peak}$-ratio mergers, and including scatter creates more major $M_{\rm star}$-ratio mergers.

Henceforth, we do not include scatter in the $M_{\rm star} - M_{\rm peak}$ relation. Thus, while the overall trends are unaffected by scatter, the merger fractions that we cite are likely \textit{lower limits}.

We also consider the difference between paired vs. isolated halos and the influence of resolution effects on our results. We find that the merger histories of satellites in the paired and isolated halos are not significantly different. For example, at $M_{\rm star} > 10 ^ 6 M_\odot$, only a slightly higher fraction (10\%) of satellites experienced a major merger since $z = 1$ in the paired halos relative to the isolated halos. We also find that the high-resolution runs tend to have mergers $\sim 1-2$ Gyr earlier than the fiducial runs. However, this simply tweaks the timescale of the merger, as opposed to the overall number. In any case, our definition of (recent) merger timescales are somewhat arbitrary (as discussed in \S \ref{sec:times}), so this does not change our overall conclusions.

\subsection{Mergers Before and After Virial Infall}
\label{sec:inf}

In the previous sub-section, we found that a non-negligible fraction of dwarf galaxies experienced a major merger since $z = 1$. Particularly for those that are inside the host virial radius today, it is instructive to consider \textit{when} and \textit{where} these mergers happened.

\begin{figure}
\begin{center}
\includegraphics[width=8.5cm, height=6.8cm]{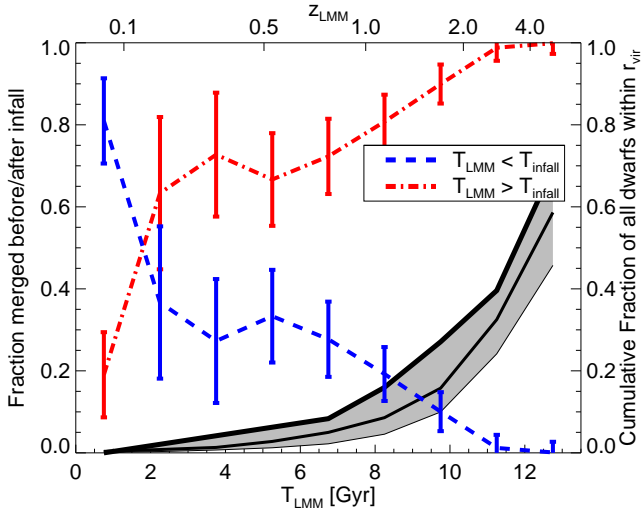}
\caption{\label{fig:fractions}
For all major mergers of dwarf galaxies with $M_{\rm star} = 10 ^ {3 - 9} M_\odot$, the fraction that occurred before (dot-dashed red) or after (dashed blue) first infall into the MW/M31-like host halo as a function of time of last major merger, $T_{\rm LMM}$. For context, the black solid curves in the gray shaded region indicate the cumulative fraction of all satellites within $r_{\rm vir}$ that experienced a major merger as a function of $T_{\rm LMM}$, in the three mass bins from Fig.~\ref{fig:tmm} (thicker curves show more massive satellites). Approximately, 3 - 10\% of all satellite dwarf galaxies experienced a major merger since $z = 1$, depending on mass (see Fig.~\ref{fig:tmm} and Table~\ref{tab:fractions}). Most major mergers in the histories of current satellites occurred before virial infall, but recent major mergers ($T_{\rm LMM} < 2$ Gyr) more likely occurred after infall.}
\end{center}
\end{figure}

Fig.~\ref{fig:fractions} shows the relative fraction of satellite galaxies that experienced their major merger before (dot-dashed red) and after (dashed blue) falling into the host halo, as a function of the time of the last major merger. Here we define virial ``infall'' as the \textit{first} time that a satellite crossed within $r_{\rm vir}$ of the MW/M31-like host halo.

\begin{figure*}
\begin{center}
    \includegraphics[width=18cm, height=6cm]{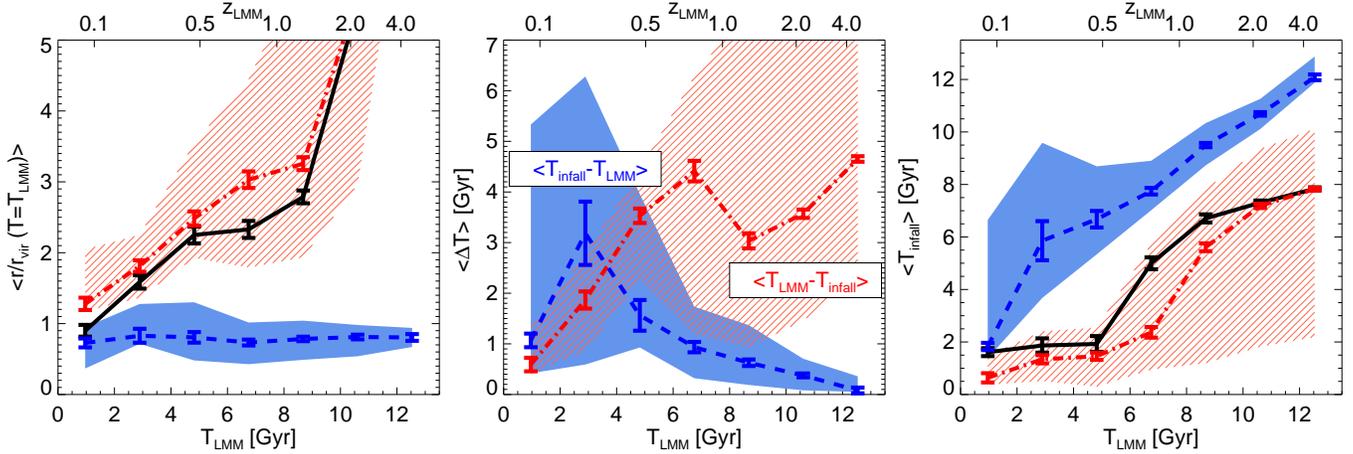}
\caption{\label{fig:rmm}\textit{Left}: The average distance, scaled by host virial radius, $r_{\rm vir}$ (at $T_{\rm LMM}$), where a satellite dwarf galaxy's last major merger occurred as a function of time since last major merger, $T_{\rm LMM}$. Error bars indicate error in the mean. The solid black curve shows the average for all mergers, while dashed blue and dot-dashed red curves are for mergers that occurred inside/outside $r_{\rm vir}$, respectively. Shaded regions indicate $1 - \sigma$ scatter. Mergers before infall at $z < 1$ typically happened at $1 - 3\,r_{\rm vir}$. Mergers after infall typically occurred \textit{just within} $r_{\rm vir}$. \textit{Middle}: The average time difference between merger and virial infall. Mergers before infall occurred $\Delta T \sim 0.5 - 5$ Gyr before crossing inside $r_{\rm vir}$, with a bigger time difference for earlier mergers. Mergers after infall typically occurred quickly, $\Delta T \lesssim 3$ Gyr after crossing inside $r_{\rm vir}$. \textit{Right:}  The average infall time for satellite dwarf galaxies as a function of time since last major merger, $T_{\rm LMM}$. More recent merger events generally have later infall times.}
\end{center}
\end{figure*}

Most major mergers occurred prior to falling into the host halo, with approximately $70 / 30 \%$ merging before/after infall since $z \sim 1$ (based on a cumulative version of Fig. \ref{fig:fractions}), in broad agreement with previous work \citep{wetzel09b}.\footnote{Though, for $\approx 20$\% of the ELVIS host halos, no major mergers of satellites occurred after infall.}
The strong gravitational tidal field inside of $r_{\rm vir}$ makes satellite-satellite mergers less likely there \citep[e.g.,][]{derijcke04}. 
However, the most recent mergers ($\lesssim 2$ Gyr ago) did occur predominantly \textit{after} infall, though $\lesssim 1 \%$ of current satellite galaxies experienced a major merger in the past $2$ Gyr.

Fig.~\ref{fig:rmm} shows more directly when and where mergers occurred, separated by whether the merger happened before or after first infall.
The left panel indicates where the merger took place relative to the host virial radius (defined at the redshift of the merger), while the middle panel shows the time difference between first infall into the host halo and the merger.
Recent mergers ($z_{\rm LMM} < 1$) that occurred prior to infall typically occurred $\sim 0.5 - 5$ Gyr before infall and at $1 - 3\,r_{\rm vir}$.
Earlier mergers that occurred before infall had larger relative distances from the host and longer time delays between mergers and infall. 
The predominance of mergers that occurred prior to infall leads to merger remnants having more recent infall times than typical subhalos (see right panel of Fig. \ref{fig:rmm}).

In contrast, mergers that occurred after infall occurred just inside the host $r_{\rm vir}$, independent of redshift, typically $\sim 1 - 3$ Gyr after infall. The evolution of this timescale roughly tracks the evolution of the halo dynamical time ($\sim 2$ Gyr at $z = 0$). The short time between infall and merging implies that such satellite-satellite mergers are driven by infalling groups, which we show explicitly in Wetzel, Deason, Garrison-Kimmel, in prep. \citep[see also][]{Angulo_2009}.

Using the Bolshoi $N$-body simulation, \citet{Behroozi_2014} recently found that major mergers rarely occur within $\sim 4\,r_{\rm vir}$ of a more massive host halo. Their result does not disagree with ours, because they examined \textit{halo-halo} major mergers near a more massive host halo, that is, one halo merging into another, creating a central-satellite (halo-subhalo) pair, a process that precedes a galaxy-galaxy merger. Indeed, if galaxy-galaxy major mergers follow halo-halo major mergers after a delay that is comparable to an orbital time from $\sim 4$ to $\sim 1\,r_{\rm vir}$ toward a more massive host halo, then this helps to explain why current satellites typically had their last galaxy-galaxy major merger near the host $r_{\rm vir}$.

\subsection{Radial Distance of Merger Remnants}

\begin{figure*}
\begin{center}
\includegraphics[width=15cm, height=10cm]{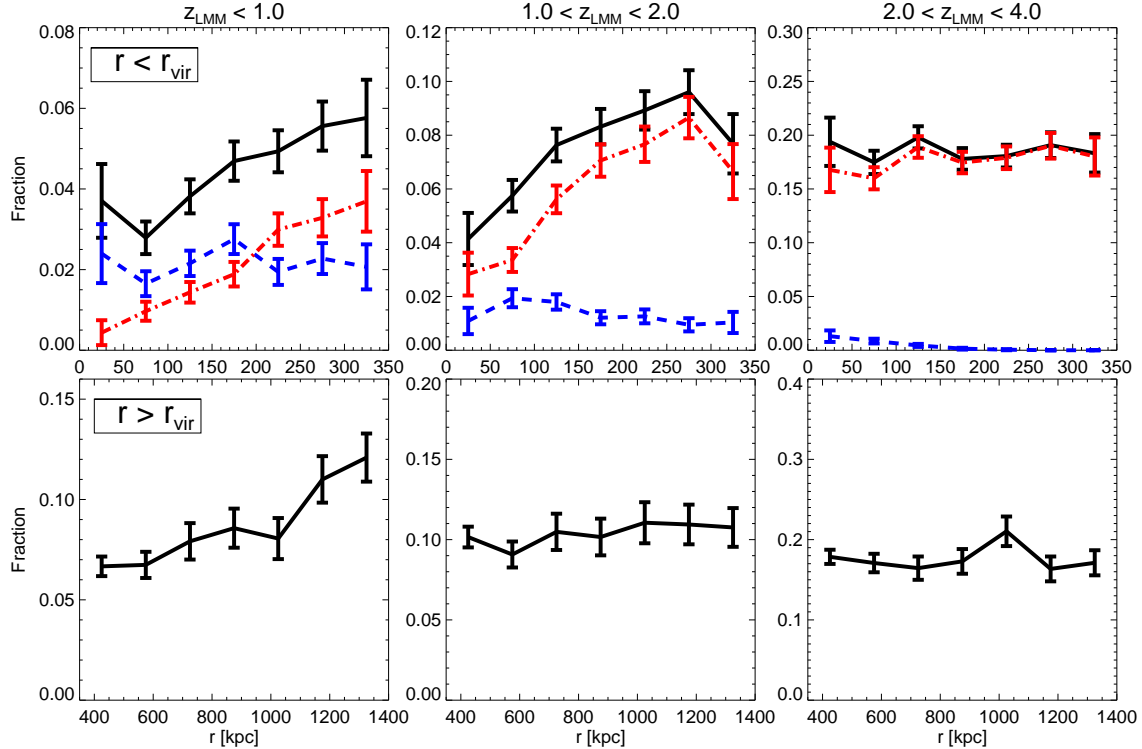}
\caption{\label{fig:rad}
\textit{Top}: Fraction of all satellite dwarf galaxies with $M_{\rm star} = 10 ^ {3 - 9} M_\odot$ currently within $r_{\rm vir}$ that experienced a major merger in the given redshift intervals (from left to right) as a function of current radial distance from the center of the host halo, $r$. The overall fraction has contributions from mergers before (dot-dashed red) and after (dashed blue) virial infall. Satellites currently at larger distances are more likely to have experienced a major merger at $z_{\rm LMM} < 2$, a trend that is driven by mergers that occurred prior to infall, which have later infall times and therefore are biased towards larger distances. \textit{Bottom}: Same as top panels but for dwarf galaxies at $r_{\rm vir} < r < 1400$ kpc. These ``field'' dwarf galaxies show similar trends as those within $r_{\rm vir}$, though with a significantly higher frequency of mergers. None of these trends with distance depend significantly on dwarf mass.}
\end{center}
\end{figure*}

We next consider the current locations of major merger remnants. Fig.~\ref{fig:rad} shows the fraction of dwarf galaxies that experienced a major merger as a function of current radial distance from the center of the (nearest) host halo. Each column shows mergers from different redshift intervals, from recent ($z_{\rm LMM} < 1$, left) to intermediate ($1 < z_{\rm LMM} < 2$, center) to early ($2 < z_{\rm LMM} < 4$, right). The figure shows all galaxies with $M_{\rm star} = 10 ^ {3 - 9} M_\odot$, though the trends do not depend significantly on mass.

The top panels show all satellites currently within the host $r_{\rm vir}$ and distinguishes between major mergers that occurred before (dot-dashed red) and after (dashed blue) virial infall. Merger remnants from $z < 2$ show a strong bias in radial distance today, such that galaxies at larger distances are more likely to have experienced a major merger. By contrast, remnants of mergers that occurred after virial infall show no bias in current distance. Thus, the overall trend is driven entirely by mergers that occurred prior to infall, because (recent) mergers that occurred prior to infall tend to have more recent infall times and therefore are biased towards larger distances. This result has important implications for future observations that will measure dwarf galaxies out to larger distances, where we expect more evidence for recent major mergers.

The bottom panels of Fig.~\ref{fig:rad} show the current distance to the nearest host halo for galaxies beyond $r_{\rm vir}$. Mergers since $z_{\rm LMM} < 1$ again are biased towards larger radial distances, but the trend disappears for remnants of earlier ($z_{\rm LMM} > 1$) mergers. Thus, we are most likely to observe the remnants of the most recent major mergers at the outer regions of the Local Group. This trend for non-satellite dwarfs with $r > r_{\rm vir}$ today is likely due to the influence of the host galaxy beyond $r_{\rm vir}$. Several works have shown that the influence of a massive host can influence subhalos out to several $r_{\rm vir}$ (e.g, \citealt{Behroozi_2014}; \citealt{wetzel14}). Thus, it is unlikely that there is a sharp change in frequency of major mergers at $r=r_{\rm vir}$.

\citet{slater13} showed that the radial distance distribution of merger remnants in the Via Lactea simulations \citep{diemand07, diemand08} are approximately flat relative to the more centrally concentrated distribution of observed dwarf galaxies in the Local Group. Our results agree with this work, because we find that (recent) merger products are biased towards larger distances \textit{relative} to the overall galaxy population.
Examining somewhat more massive systems in $N$-body simulations, \citet{wetzel09a} and \citet{Angulo_2009} found that satellite merger remnants broadly trace the overall radial distribution of satellites, in seeming disagreement with the bias to larger radial distances here.
However, they examined much more recent merger remnants ($\lesssim 200$ Myr after merging).
Indeed, our most recent satellite merger remnants, which are dominated by satellite-satellite mergers, show little radial distance bias in Fig.~\ref{fig:rad}.

\section{Implications for Dwarf Galaxies of the Local Group}
\label{sec:LG}

\subsection{Number of merger remnants in the Local Group}

There are $N \sim 46$ dwarf galaxies in The Local Group in the (likely complete) stellar mass range of $10 ^ {6 - 9} M_\odot$ \citep{mcconnachie12}. Assuming $r_{\rm vir} = 300$ kpc for the virial radii of both the MW and M31, $N \sim 16\,(30)$ reside inside (outside) $r_{\rm vir}$. Thus, the merger fractions in Fig.~\ref{fig:tmm} and Table~\ref{tab:fractions} predict that, within this mass range, both the MW and M31 halo each should contain one satellite dwarf galaxy that underwent a major merger since $z = 1$. Furthermore, approximately $N \sim 5$ of the ``field'' dwarf galaxies beyond $r_{\rm vir}$ of the MW or M31 likely experienced a major merger since $z = 1$.

Fornax is the second most luminous ($L \sim 10 ^ 7 L_\odot$) dwarf spheroidal galaxy within the MW halo, after the Sagittarius dwarf. Several studies have found that it displays complex chemical and dynamical properties \citep[e.g.,][]{battaglia06, deboer12, hendricks14}, which has led to the suggestion that Fornax underwent a recent merger \citep[e.g.,][]{coleman04, yozin12, amorisco12}. Similarly, the Carina dwarf spheroidal galaxy ($L \sim 10 ^ 6 L_\odot$) also displays chemical peculiarities \citep[e.g.,][]{venn12} that may indicate recent merger activity. Furthermore, the apparent kinematic twist of And II \citep{ho12}, a satellite of M31 with mass similar to Fornax, led \citet{amorisco14} to suggest that it is the remnant of a dwarf-dwarf major merger \citep[see also][]{lokas14}.

In the context of our results, these suggestions are plausible; specifically, $\sim 1 - 3$ satellite dwarf galaxies in the halos of the MW and M31 probably experienced a major merger since $z = 1$. Although dwarf-dwarf mergers are not particularly common, neither are they extremely rare. Thus, our results highlight the importance of considering mergers in interpreting the star formation histories, metallicity distribution functions, and chemical abundance patterns of dwarf galaxies, particularly for systems such as Fornax or And II which display kinematic irregularities.

Finally, we find that major mergers are even more important for ``field'' dwarf galaxies in the Local Group: a significant fraction ($\sim 15 - 20\%$) at $M_{\rm star} > 10 ^ {6} M_\odot $ likely experienced a major merger since $z \sim 1$. Thus, major mergers are more common farther away from massive host halos, a result that contradicts any expectation that this is a regime where galaxy interactions are unlikely. Indeed, one should not label the most distant dwarf galaxies of the Local Group as necessarily ``isolated'' because of the lack of galaxy companions; such dwarf galaxies may well have had a recent merger, but the companion is no longer observable (in agreement with the results of \citealt{Hirschmann_2013} for more massive ``isolated'' galaxies). Thus, we stress the importance of considering mergers in interpreting observations even of apparently isolated dwarf galaxies in the Local Group.

\subsection{Gas content of mergers}

The distinction between mergers that occurred before or after virial infall may be important in understanding the gas content of mergers. Empirically, almost all dwarf galaxies at distances $> r_{\rm vir}$ of the MW/M31 halo (limited to $M_{\rm star} \gtrsim 10 ^ 6 M_\odot$) are gas-rich and star-forming \citep[e.g.][]{grcevich09, geha12}. Therefore, dwarf-dwarf mergers beyond the host $r_{\rm vir}$ are likely gas-rich. On the other hand, almost all satellite dwarf galaxies inside $r_{\rm vir}$ of the MW/M31 are gas-poor and quenched, likely because of strong tidal/ram-pressure stripping by the host halo. Therefore, satellite-satellite mergers after infall are more likely to be gas-poor dissipationless mergers.

The effects on a galaxy from a gas-rich versus gas-poor merger is likely quite different, though a quantitative understanding for the Local Group remains difficult, given that the timescales for gas stripping after virial infall, the effects of group pre-processing, and the influence of the host halo at several $r_{\rm vir}$ remain poorly known. Satellite-satellite mergers may in fact remain gas-rich if the short time delay between virial infall and merging ($\Delta T \sim 1 - 3$ Gyr) is shorter than the environmental quenching timescale after virial infall, which can be long (e.g. \citealt{wetzel13}; \citealt{wheeler14}), at least for more massive satellites. On the other hand, the near-unity quenched fraction of satellites of the MW and M31 suggests much faster quenching for dwarf satellites.
While it is beyond the scope of this work (based on simulations with only dark matter) to quantify the gas content of mergers, it would be an interesting extension to examine gas-rich versus gas-poor dwarf-dwarf mergers in hydrodynamical simulations.

\subsection{Observability of merger remnants}
\label{sec:times}
Throughout, we labeled ``recent'' mergers as those that occurred after $z = 1$.
This redshift does have some physical significance in the evolution of dwarf galaxies. For example, \citet{weisz11} showed that most dwarf galaxies formed the bulk ($\sim 60\%$) of their stellar mass by $z = 1$. However, we acknowledge that the exact choice is somewhat arbitrary, because it remains unclear how long major mergers would have significant remaining impact on star formation, stellar kinematics, or morphology, and indeed it may be different for each of these quantities. Furthermore, the survival of merger signatures in phase-space depends strongly on the gravitational potential as well as the merger orbit. We plan to investigate more observationally motivated timescales for dwarf-dwarf mergers in future work.

One area in which all major mergers, regardless of redshift, are relevant is archaeologically derived histories of star formation and metallicity for dwarf galaxies from resolved stellar populations \citep{weisz14}.  Any boosts in star formation may well be the by-product of gas-rich mergers, or similarly, mergers between galaxies of discrepant stellar mass could lead to significant scatter in the internal metallicity distribution. Our results suggest that such observational signatures are most likely in the history at earlier times, when mergers were most likely, and for dwarf galaxies at larger distances from the center of the MW/M31.

\subsection{Massive Black Holes in Dwarf Galaxies}

Major mergers may have important implications for massive black holes at the centers of dwarf galaxies. In recent years, several authors have pursued a concerted effort to study massive black holes in dwarf galaxies \citep[e.g.,][]{reines13}, motivated in part because these mass scales are likely the most closely linked to primordial seeds of massive black holes in low-mass galaxies. However, our results suggest that major merger activity is non-negligible on dwarf mass scales. Such galaxy major mergers may lead to the merging of the constituent massive black holes, and gas-rich mergers (which are likely for dwarf galaxies) could trigger rapid gas accretion into the black hole, both of which would enhance the growth of the central black hole beyond the primordial seed. Thus, dwarf-dwarf mergers may play an important role in our ability to distinguish between formation scenarios of massive black hole seeds using local dwarf galaxies.

Furthermore, a merger between dwarf galaxies that both contain a massive black hole can lead to a black hole merger with a gravitational-wave recoil \citep{bekenstein73, fitchett84}. While the recoil velocity depends on the mass ratio and relative spin of the black holes, the lower escape velocity of dwarf galaxies (relative to more massive galaxies) means that recoiling black holes more easily can escape from the galaxy. Thus, black hole ejection may be an important consequence of mergers between dwarf galaxies. In particular, if all dwarf galaxies host massive black holes, then those dwarfs that experienced a major merger may no longer retain a black hole. Thus, in order to quantify the occupation distribution of black holes in nearby dwarf galaxies, one should consider the major merger fractions that we present here.

\section{Conclusions}

Dwarf galaxies in the Local Group are excellent laboratories to study dark matter, gas dynamics, star formation, and chemical evolution on the smallest cosmological scales. Such properties of dwarf galaxies are affected by external environmental processes, and dwarf-dwarf galaxy mergers are one such process that can have a strong influence. We investigated the incidence of dwarf-dwarf galaxy mergers in the Local Group using the ELVIS suite of cosmological zoom-in dissipationless simulations. Our main conclusions are as follows.

\begin{enumerate}

\item Recent dwarf-dwarf galaxy major mergers are not uncommon. At $M_{\rm star} > 10 ^ {6} M_\odot$, 10\% of satellite dwarf galaxies within the virial radius of a MW or M31-like host halo experienced a major merger since $z = 1$. The incidence of mergers becomes less frequent for lower mass dwarf galaxies. 

\item For satellites inside of the host virial radius today, most mergers occurred \textit{before} virial infall. The satellite-satellite mergers that occurred after infall typically happened shortly ($\sim 1 - 3$ Gyr) after crossing within the host virial radius and were between galaxies that fell into the host halo as a pair in a group (see Wetzel, Deason, Garrison-Kimmel 2014, in prep. for more details).

\item Satellite merger remnants are biased towards larger radial distances within the host halo, because most mergers occurred prior to falling into the host halo, and thus merger remnants have more recent virial infall times.

\item Dwarf galaxies that are beyond the host virial radius are about twice as likely to have experienced a recent merger. Specifically, $15 - 20\%$ of dwarf galaxies with $M_{\rm star} > 10 ^ 6 M_\odot$ at $r_{\rm vir} < r < 1400$ kpc experienced a major merger since $z = 1$. Such dwarf merger remnants also are more likely at larger distances from the host halo. Thus, for dwarf galaxies at large distances, where interactions with a large halo such as MW or M31 are less likely, dwarf-dwarf mergers are \textit{more} likely.

\end{enumerate}

\section*{Acknowledgments}
AJD is currently supported by NASA through Hubble Fellowship grant HST-HF-51302.01, awarded by the Space Telescope Science Institute, which is operated by the Association of Universities for Research in Astronomy, Inc., for NASA, under contract NAS5-26555. We thank the Aspen Center for Physics and National Science Foundation Grant \#1066293 for hospitality during the writing of this paper. We thank Nicola Amorisco, Vasily Belokurov, Gurtina Besla, Laura Blecha, and Wyn Evans for useful discussion. We also thank the anonymous referee for useful comments.

\end{document}